\documentclass[11pt,twoside]{article}
\usepackage{asp2010}
\resetcounters

\begin{document}

\title{The magneto-ionic medium in the Milky Way}
\author{Marijke Haverkorn}
\affil{$^1$ASTRON, PO Box 2 7990 AA Dwingeloo, Netherlands}
\affil{$^2$Leiden Observatory, Leiden University}

\begin{abstract}

One way in which the Canadian Galactic Plane Survey has made an
important contribution to the understanding of the Galactic
interstellar medium is through its polarization surveys. Investigation
of these data has enabled a big step in the study of magnetic fields
in the interstellar medium and a range of discrete, extended,
interstellar objects. In this review, I will discuss the role that the
magnetic field plays in the interstellar medium, summarizing the ways
in which magnetic field interacts with the other components in the
Milky Way. Magnetic fields in the Galactic halo are discussed, and an
outlook to a number of successor surveys of the polarized CGPS in the
near future is given.

\end{abstract}

\section{Introduction}

The Canadian Galactic Plane Survey (CGPS) has had broad and far
reaching consequences. In particular the radio polarimetric continuum
survey at 1420 MHz greatly influenced the study of magnetic fields in
the Milky Way. This is largely due to the careful analysis of a
variety of magnetic structures such as polarization lenses
\citep{gld98, ul02}, Galactic chimneys \citep{wen07}, supernova
remnants \citep[e.g.][]{ukb02, f05, kff06}, pulsar wind nebulae
\citep{kru06, klr08, ruk08}, H~{\sc II} regions \citep{gld99, fks06},
magnetic reversals in spiral arms \citep[][Van Eck et al, this Volume;
Rae et al, this Volume]{browntaylor01,btw03}, culminating in the
overview of the entire Canadian Galactic Plane in polarization
\citep{lrr10}.

In this contribution to the meeting celebrating the CGPS, I will
discuss the role that magnetic fields play in the interstellar medium
(ISM) of the Milky Way. In Section~\ref{s:energydensities}, I will
discuss energy densities of the various components in the Galactic
ISM, to show the dynamic importance of the magnetic field in the Milky
Way. Section~\ref{s:role} focuses on the various effects that the
magnetic field has on the Galactic ISM, including new CGPS results
concerning interstellar turbulence. A discussion of the magnetic field
in the halo of the Milky Way can be found in
Section~\ref{s:halo}. Finally, in Section~\ref{s:future} we provide an
outlook to the (bright!) future of radio polarimetry for studying
galactic magnetism.

\section{Energy densities in the Galactic interstellar medium}
\label{s:energydensities}

The strength of the influence of magnetic fields on the ISM is
commonly quantified by comparing energy densities or pressures of the
various components. The magnetic field energy density is found to be
comparable to the cosmic ray energy density, while the kinetic energy
density (thermal and turbulent cold and warm gas pressures) can be in
equipartition or up to twice as high as the equipartition
value \citet[e.g.][]{bc90}. Thermal energy densities are about a
magnitude lower.

In nearby external galaxy NGC~6946, \citet{b04} estimated the energy
densities of the magnetic field from synchrotron emission, of the
thermal warm ionized gas from thermal radio emission maps, of the
thermal neutral (molecular and atomic) gas from CO and HI maps, and of
the turbulent component from HI line widths. He obtained the left hand
plot in Figure~\ref{f:energydens}, which shows energy densities of
these various components of the magneto-ionic medium in the galaxy
NGC~6946 as a function of galactocentric radius. \citet{b04} concluded
that the magnetic field energy density is comparable to the turbulent
gas density in the inner parts of NGC~6946, and dominates in the outer
galaxy. Both components are more than a magnitude stronger than the
thermal gas energy densities. The energy density of the rotation of
the neutral HI gas is about 500 times higher than the turbulent gas
density. However, the filling factor of the neutral gas is assumed to
be 1 in these calculations, which may result in an overestimate of the
energy density of the turbulent gas.

%%This is also discussed in Fletcher (this volume).

%Stepanov et al (2009): inconsistent with local equipartition as CRs
%  smooth on large scales and Bfld turbulent on small scales

The right hand side of Figure~\ref{f:energydens} shows a similar plot
for the Milky Way. The magnetic field energy density $|B|^2/8\pi$ is
based on equipartition values derived from radial profiles as modeled
by \citet{bkb85}. This method was used by Elly Berkhuijsen in
\citet{b01} to obtain the Galactic magnetic field strength as a
function of Galactocentric radius. The figure also shows the
difference between the classical equipartition formula and the revised
formula based on integration over a fixed energy range instead of a
frequency range \citep{bk05}. The thermal gas energy density is only
indicated at the solar radius and is based on standard values of the
densities and temperatures of the cold, warm and hot gas components at
the solar radius. The turbulent gas energy density $\rho v_{turb}^2$
is calculated using an exponential gas scale length of 3.15~kpc and a
turbulent velocity based on \citep{md07}. They found that the
turbulent velocity in the (inner) Galaxy is best described by three
components: a cold component with a velocity dispersion $\Delta v =
6.3$~\mbox{km~s$^{-1}$}, a warm component with $\Delta v =
12.3$~\mbox{km~s$^{-1}$}, and a fast component of $\Delta v =
25.9$~\mbox{km~s$^{-1}$}, probably related to large-scale motions. All
components are more or less constant with radius. Both the warm
turbulent component with $\Delta v = 12.3$~\mbox{km~s$^{-1}$} and the
large-scale motion component with $\Delta v = 25.9$~\mbox{km~s$^{-1}$}
are shown in the figure. Throughout, a solar radius of 8.5~kpc is used.

Qualitatively, the situation in the Milky Way is similar to NGC6~946:
the rotational energy density is more than two orders of magnitude
higher than the energy densities of all other components. In addition,
the magnetic and turbulent energy densities are, given the
uncertainties in the assumptions made, not far off from each other,
while the energy density in the thermal gas components is much
less. The dominance of magnetic pressure over the thermal pressure is
consistent with the observed remarkable uniformity in magnetic field
strength over a wide range of gas densities \citep{th86}.  This
reconfirms what we knew already: that the magnetic field in the Galaxy
is a major player, in any case on large scales, which cannot be
ignored.

\articlefiguretwo{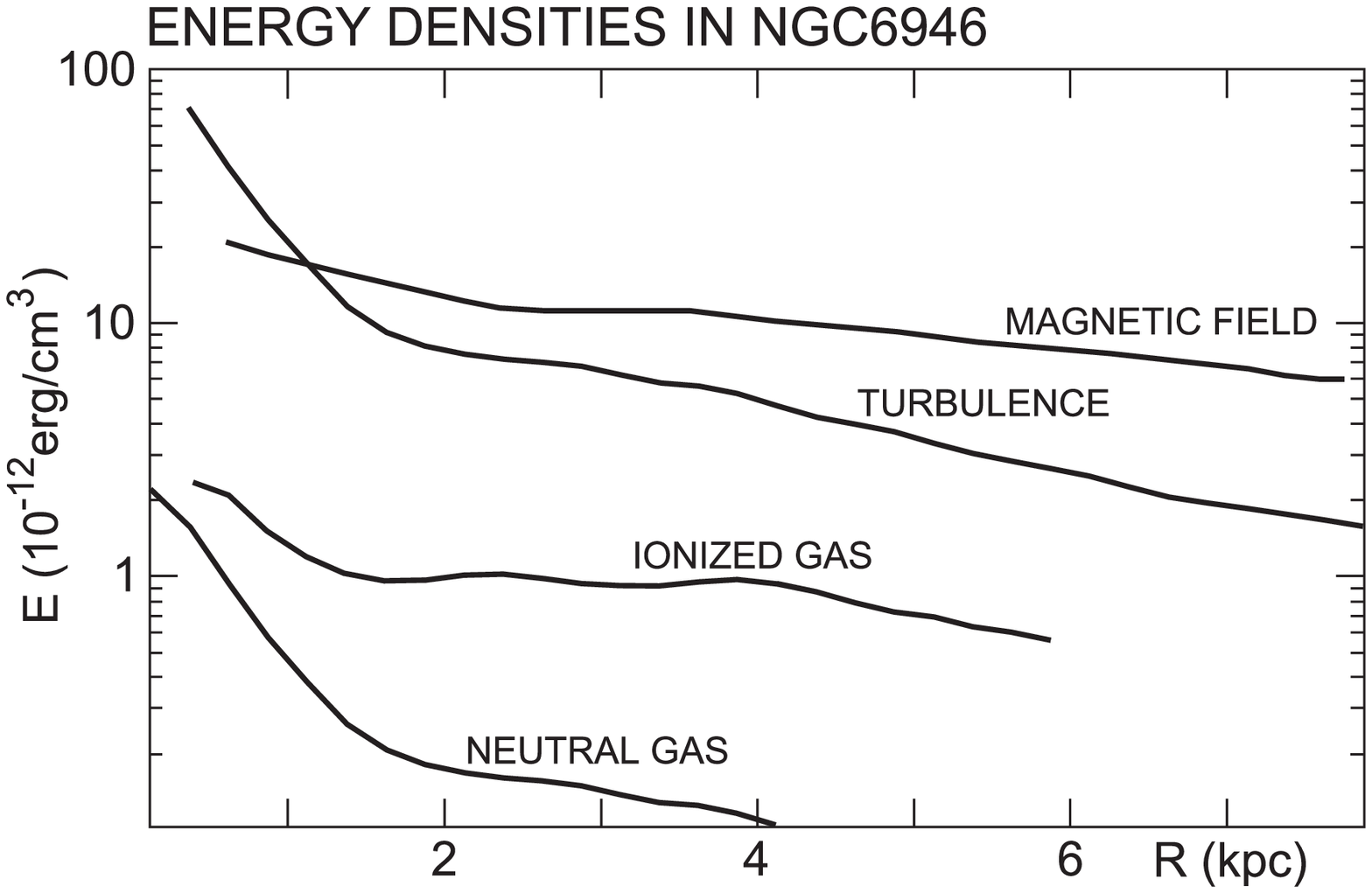}{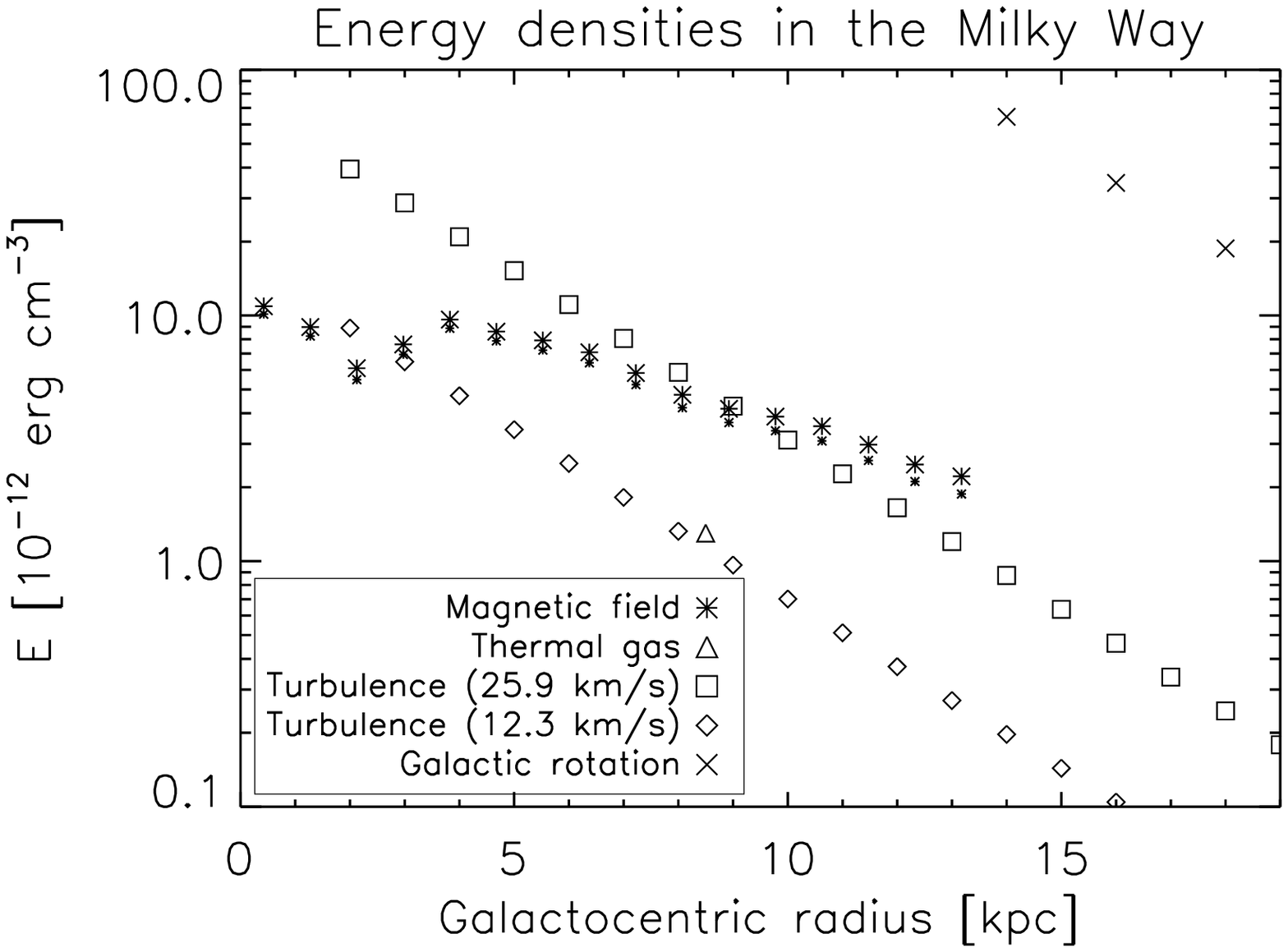}{f:energydens}{Left:
Energy densities of various components of the interstellar medium in
NGC~6946 \citep{b04}. Image courtesy of Rainer Beck, reproduced by the
kind permission of Springer Publishers. Right: Galactic ISM as a
function of Galactocentric radius: the magnetic field according to the
equipartition function revised by \citet{bk05} (large asterisks),
magnetic field according to the classical equipartition formula (small
asterisks), gas with a velocity dispersion of 25.9~km~s$^{-1}$ (boxes)
and of 12.3~km~s$^{-1}$ (diamonds), thermal gas at the solar radius
(triangle) and Galactic rotation (crosses).}

%cosmic ray energy density assumed in equip to calculate Bfield?

%pressure vs energy density?

%Check CRS two components -- gammarays two components? If not, then Bfield two components.

%*** See in Ferriere 2001 p1053 for derivation of Bmag from CR pressure

\section{The role of the magnetic field in the Milky Way}
\label{s:role}

Magnetic fields interact with charged particles through the Lorentz
force, which makes the particles gyrate around magnetic field lines.
As most of the Universe is ionized, magnetic fields have a major
influence on many of the physical processes in the Universe (mostly
those where gravity is not important). Neutral particles are coupled
through ion-neutral collisions. Therefore, even with fractional
ionizations of $10^{-6}$ to $10^{-8}$ typically found in dense cloud
cores \citep{cwt98} the neutral interstellar medium is sufficiently
ionized to expect a significant connection to magnetic fields.

There is a myriad of consequences of the interact of Galactic magnetic
fields and the interstellar gas. Below I try to summarize the most
important one in our Milky Way.

\paragraph{Additional pressure component}

The magnetic field in the Milky Way provides a significant pressure
component, comparable to the turbulent gas and cosmic ray
pressures. This pressure component contributes significantly to the
total pressure which counterbalances the thick gas disk ($\le 1$~kpc)
against gravity \citep{bc90}. This component also causes slower
expansion of supernova remnants, see below.

%CR pressure would be 1/3 of energy density if all CRs were
%ultra-relativistic. However, the bulk of CR energy density is due to
%mildly relativistic protons \citep{bc90} which makes CR energy density
%a little higher. Applying to observations of cCR pressure just outside
%the heliosphere gives a value of $P_{CR} \approx
%12.8\times10^{-13}$~dyn~cm^{-2}.

\paragraph{Magnetic fields and spiral arms}

Magnetic fields are also expected to influence the way in which gas
flows through the spiral arms. \citet{gc02} discuss numerical
magneto-hydrodynamical simulations of gas flows in a Galactic
potential with a spiral perturbation. They show that simulations
including magnetic fields show increased vertical velocity structure
of the gas falling into the spiral arms, and a clumpier density
distribution, than hydro-dynamical simulations. Large-scale magnetic
field reversals along spiral arms could play a role in the gas
dynamics through current sheets and/or increased magnetic
reconnection, but there is still much debate about the number and
location of reversals in the Galactic magnetic field (see e.g.\ Brown,
this volume). \citet{s05} argues that the only unequivocally observed
large-scale reversal in the Galactic magnetic field could in fact be a
'localized' feature of a few kpc in size instead of a full-fledged
magnetic field reversal along a spiral arm.

\paragraph{Cosmic Ray propagation and acceleration}

Cosmic rays are coupled to interstellar magnetic fields in two ways.
As cosmic rays are charged, they gyrate around magnetic field
lines. The streaming motions of these particles along the field lines
excite resonant Alfv\'en waves which then scatter the cosmic rays
\citep[e.g.][]{kp69}. It is this scatter of cosmic rays off magnetic
field irregularities that accelerates them to their relativistic
velocities. This can happen as first-order Fermi acceleration in
expanding supernova shock fronts \citep{bo78}, or stochastically
through second-order Fermi acceleration in a turbulent gas
\citep{f49}.  

Each cosmic ray electron or ion is primarily sensitive to magnetic
field fluctuations on the same scale as its Larmor radius. Therefore,
low-energy cosmic rays (below about $10^{15}$~eV) are effectively
scattered by the turbulent magnetic field, which makes direct tracing
back to their origins impossible. It also confines the cosmic rays to
the Galaxy. High-energy and ultra-high-energy cosmic rays are mostly
affected by the uniform field component. Measurements of arrival
directions of ultra-high energy cosmic rays by the Pierre Auger
Observatory show possible clustering of detections around a number of
nearby active galactic nuclei, among which Centaurus~A
\citep{auger07}. As the deflection by the Galactic magnetic field is
believed to be a few to a few tens of degrees, depending on magnetic
field strength and direction and on cosmic ray composition, accurate
measurements/modeling of the large-scale Galactic magnetic field is
also vital for understanding the origins of ultra-high-energy cosmic
rays.

It is this scatter of cosmic rays off magnetic field irregularities
that accelerates cosmic rays to their relativistic velocities. This
can happen as first-order Fermi acceleration in expanding supernova
shock fronts \citep{bo78}, or stochastically through second-order
Fermi acceleration in a turbulent gas \citep{f49}.

%-- Prouza \& Smida 

%Strong review says:
%82. Alvarez-Mu niz J, Engel R, Stanev T. Astrophys. J. 572:185 (2002) 
%83. Takami H, Yoshiguchi H, Sato K. Astrophys. J. 639:803 (2006) 
%84. Kachelriess M, Serpico PD, Teshima M. astro-ph/0510444 (2005) 

\paragraph{Supernova remnants and superbubbles}

%In Ferriere 2001: Tomisaka (1990), Ferriere et al 1991 (fmz91);
%Slavin \& Cox 1992

Surrounding large-scale magnetic fields have a direct impact on
expansion of supernova remnants (SNRs). Magnetic pressure from the
surrounding magnetic field slows the expansion of a SNR \citep{fmz91},
while increased magnetic tension perpendicular to the magnetic field
lines causes anisotropy in the expansion in the direction of the
surrounding magnetic field. Also, magnetic fields limit the
compression of gas in the shock (as the magnetic pressure increases as
the square of compressed magnetic field), which results in smaller
SNRs with thicker shells \citep{fmz91} and a smaller filling factor of
the hot ionized medium than a situation without magnetic
fields. However, the opposite effect is reached by significant
large-scale azimuthal magnetic fields in superbubbles, which can
prevent bubbles from breaking out of the disk, allowing them to grow
larger.

 \citet{swo09} performed magneto-hydrodynamical simulations of
superbubble explosions in a magnetized ISM. Comparing their
simulations to Galactic plane surveys, among which the CGPS, they
concluded that calculated estimates of both ages of superbubbles and
the scale height of the medium that they propagate in should be
corrected by a factor 2 to 4 when including magnetic fields in the
analysis.

Magnetic fields can have opposing effects the disk-halo interaction. A
magnetic field parallel to the Galactic plane can oppose break-out of
the gas \citep{ni89}. On the other hand, when gas does break out and
magnetic field lines open up into the Galactic halo, it provides a
funnel through which charged particles can easily escape the Galactic
disk. \citet{wen07} discuss multi-wavelength observations of a
fragmenting superbubble associated with the H~{\sc II} region W4 and
find slightly enhanced magnetic fields in the shell wall of which the
component parallel to the line of sight is $\sim 3-5~\mu$G.

%- effective sound speed of the gas is increased by the presence of
%  strong fields which reduce the shock strength (Beck in LNP
%  Wielebinski \& Beck)

\paragraph{Interstellar turbulence}

Turbulence is a very important effect in the interstellar medium as it
redistributes energy from supernova explosions back into the ISM on a
wide range of scales, and it maintains the Galactic magnetic field
through (small-scale) dynamo processes. In addition,
magneto-hydrodynamical turbulence is a significant source of heating
for the ISM \citep[e.g.][]{se04}.

%Simulations show clearly the differences between hydrodynamic and
%magneto-hydrodynamic turbulence (REF). Schekochihin... find that
%magnetic energy is mostly present on much smaller scales than the gas
%energy....

Turbulence is most readily observed and studied through observations
of the 21cm neutral hydrogen line, which give direct information about
the velocity field but are not suited to study the turbulent magnetic
field. The turbulent component of the magnetic field in the ionized
gas is investigated using polarized synchrotron emission, the
Chandrasekhar-Fermi effect, but primarily Faraday
rotation. \citet{hbg08} determined from Faraday rotation of
extragalactic sources in the Southern Galactic Plane Survey
\citep[SGPS][]{hgm06} that turbulent properties in the spiral arms and
in interarm regions were distinctly different. Their Fig.~2,
reproduced here as Fig~\ref{f:sf_sgps}, shows structure
functions\footnote{The (second order) structure function of a function
$f$ is defined as $D_f(\delta\theta) =
\langle(f(\theta)-f(\theta+\delta\theta))^2\rangle_{\theta}$, where
$\theta$ is the position of a source in angular coordinates,
$\delta\theta$ is the separation between sources, i.e.\ the scale of
the measured fluctuation, and $\langle\rangle_{\theta}$ means
averaging over all positions $\theta$.} of rotation measure (RM) for
lines of sight primarily going through interarm regions (top row) and
through spiral arms (bottom row). They concluded that RM fluctuations
are present in interarm regions up to scales of about 100~pc, while in
the spiral arms, no RM fluctuations on scales larger than a few
parsecs exist.

Structure functions of RM in the CGPS region, for different directions
of the line of sight, are shown in Fig.~\ref{f:sf_cgps}. The structure
function is seen to be almost flat in the direction of the outer
Galaxy, where the Perseus spiral arm is relatively nearby, while
structure functions get steeper when going towards lower Galactic
longitude, where the Perseus arm is located further away.

\articlefigure{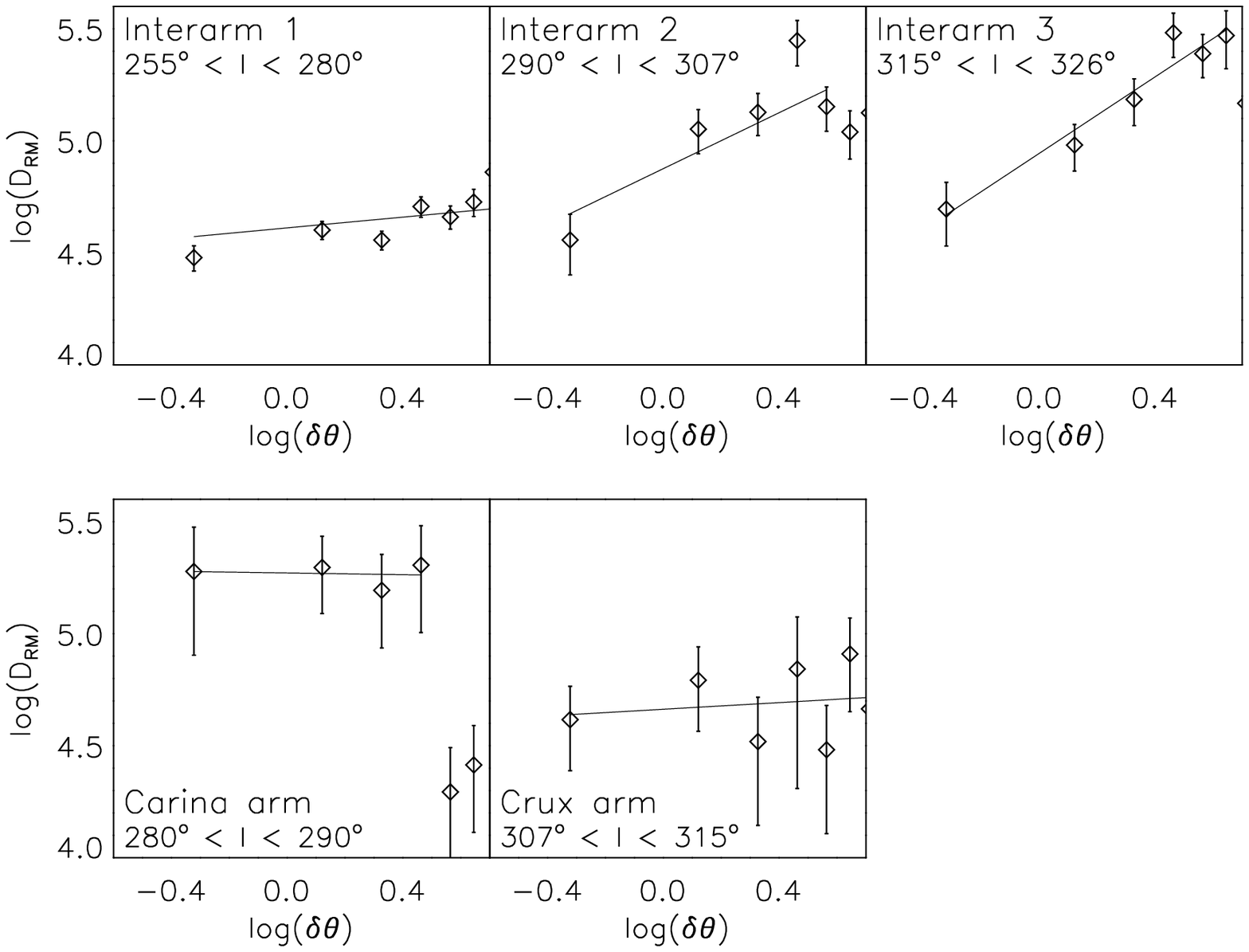}{f:sf_sgps}{Structure
functions of rotation measure D$_{\mbox{RM}}$ as a function of angular
scale $\delta\theta$ for lines of sight in the SGPS primarily through
interarm regions (top row) and lines of sight mainly crossing spiral
arms (bottom row) \citep{hbg08}. Reproduced by the kind permission of
the AAS.}

The same separation into spiral arm and interarm regions with
polarized extragalactic point sources in the Canadian Galactic Plane
Survey \citep{btw03} is not possible because in all CGPS sight lines,
spiral arms and interarm regions are superposed. However, we can
assume that the conclusions drawn from the SGPS for the inner Galaxy
are valid for the outer Galaxy as well and calculate the expected
structure functions for a superposition of spiral arms and interarm
regions. We took a very simple modeled structure function with
contributions from the Perseus arm and from the interarm region in
front of the Perseus arm: D$_{\mbox{\tiny RM}}=$D$_{\mbox{\tiny
RM,arm}}+$D$_{\mbox{\tiny RM,int}}$. D$_{\mbox{\tiny RM,arm}}$ has an
outer scale of 2~pc and a maximum amplitude of $2\sigma_{\mbox{\tiny
RM}}=100$~rad~m$^{-2}$, and D$_{\mbox{\tiny RM,arm}}$ has an outer
scale of 150~pc and a maximum amplitude of $2\sigma_{\mbox{\tiny
RM}}=250$~rad~m$^{-2}$. Since the Perseus arm is located at a
different angular scale for the three plotted regions, the
superposition of the two components is different. Fig.~\ref{f:sf_cgps}
clearly shows that the hypothesis of a small outer scale of
fluctuations in the spiral arms, as opposed to the outer scale in the
interarm regions, can also explain the observed RM structure function
in the CGPS.

\articlefigure{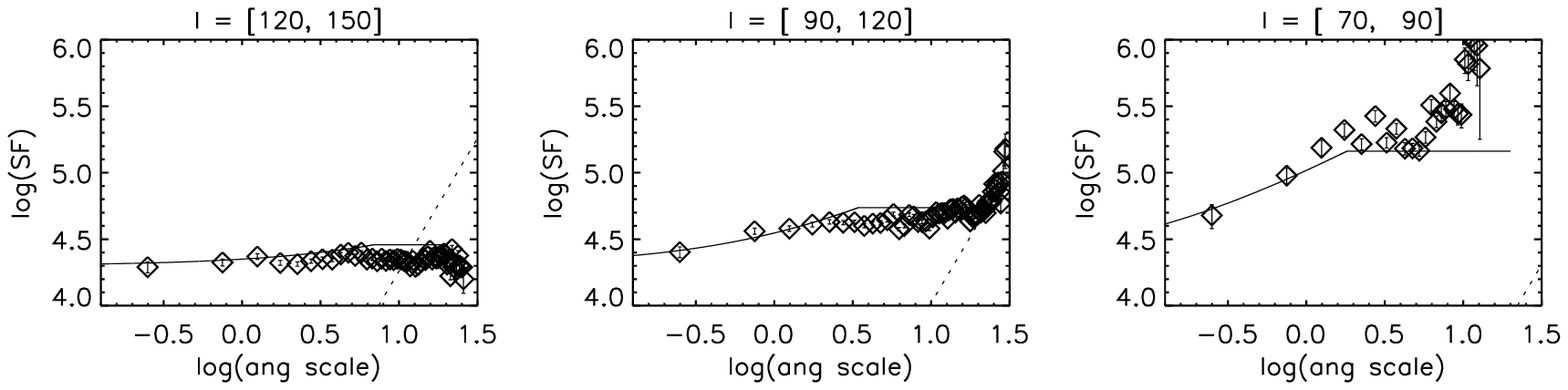}{f:sf_cgps}{Structure
functions of rotation measure D$_{\mbox{RM}}$ as a function of angular
scale $\delta\theta$ for lines of sight in the CGPS. The overplotted
lines are modeled structure functions, see the text.}

%Bflds affect dynamics in gas clouds (Elmegreen 1981, de Av\&Br 2004). 

\paragraph{Star formation}

That magnetic fields play a role in some stages of star formation is
clear, but how big that role is in which stages is still under heavy
debate. The 'classical' theory of star formation describes star
formation as a quasi-static process where collapse is slowed down or
prevented by magnetic fields \citep{sal87}. This means that magnetic
fields have to be strong enough to counter self-gravity, which is
expressed as a mass to magnetic flux ratio below a certain value
(sub-critical) \citep{ms76}.

Observations of Zeeman splitting in cold cores, however, typically
measure magnetic fields large enough to suggest that most cold clouds
are magnetically super-critical \citep[e.g.][]{cet75, tc08}. The
implication is that dense cores are fairly strongly magnetized. This
could happen if dense cores form out of magnetically sub-critical
clouds through ambipolar diffusion\footnote{Ambipolar diffusion is the
process in which charged particles are frozen in to the magnetic
fields but neutral particles can drift under influence of gravity,
slowed down by frictional drag from the ions.}  \citep{ls89}.

The alternative idea is the 'turbulent' model, where turbulence is the
main mechanism controlling star formation rather than magnetic fields
\citep[e.g.][]{e00,osg01}. Even though the magnetic field may not be
strong enough to prevent global collapse, it can still play an
important role in providing a source of pressure \citep[e.g.][]{pb07}.

\paragraph{Magnetic reconnection}

Magnetic reconnection occurs when magnetic field lines reorder
themselves into a configuration of lower energy. This has two
consequences: a change in topology of the magnetic field, and the
release of magnetic energy into motion and/or heat.

Localized magnetic reconnection in the warm ionized medium at high
latitudes has been invoked as a way to ionize the high-latitude gas in
the Galactic halo \citep{bln98}. \citet{zlb97} suggested that magnetic
reconnection due to interaction of high-velocity clouds with the
interstellar medium may be a source of heating of the interstellar
gas. Fast reconnection on small scales in a magnetized turbulent
medium could considerably increase the reconnection rate and allow
efficient mixing of magnetic fields in the direction perpendicular to
the local magnetic field direction \citep{lc04}. Fast reconnection
could avoid $\alpha$-quenching in the $\alpha-\omega$-dynamo
\citep{lv99}, and carry away angular momentum from molecular cores in
the process of star formation \citep{lsg10}.

\section{Magnetic fields in the Milky Way halo}
\label{s:halo}

Magnetic fields reside not only in the Galactic gaseous disk, but also
in the Galactic thick disk, or halo, magnetic field seems to be
decreasing very little at least out to a few kpc. An estimated
magnetic field scale height of about 4.5~kpc is a direct consequence
of measured/modeled scale heights of synchrotron emission
\citep{bkb85}. From high-latitude pulsar rotation measures,
\citet{hq94} derive a magnetic field scale height of $1.2 \pm
0.4$~kpc, a little higher than the estimate of 0.8~kpc by
\citet{sk80}. These values are reconciled if the Galactic magnetic
field becomes less regular away from the Galactic plane \citep{bc90},
or if one assumes two magnetic layers with different scale heights and
properties \citep{hq94}, consistent with the two-layer model for
synchrotron emission by \citet{bkb85}.

It is generally believed that galactic magnetic fields are maintained
and amplified by some kind of dynamo mechanism
\citep[e.g.][]{rss88,s02,w02}). The simplest model is that of the
$\alpha-\omega$ dynamo, which amplifies the radial magnetic field
component through differential rotation, and amplifies the azimuthal
and poloidal components of the field by turbulent loops twisted by the
Coriolis force. Although the $\alpha-\omega$ dynamo is believed to act
in the Sun \citep{o03}, it is considered unable to sufficiently amplify galactic
magnetic fields to observed values (eddy diffusion time scales are
much shorter than time scales for amplification of the regular field,
thereby suppressing turbulent motions, and quenching the
dynamo). Various solutions to this problem are discussed in
\citet{w02}.

For a flat disk-like galaxy which is differentially rotating,
mean-field dynamo theory predicts a quadrupolar magnetic field
configuration (the left side of Fig.~\ref{f:widrow}), where the
direction of the azimuthal magnetic field is the same above and below
the plane, but the direction of the vertical magnetic field component
reverses with respect to the plane. However, for a spherical, weakly
rotating, structure - such as possibly a Galactic halo - the dipolar
configuration is more easily excited, i.e.\ an azimuthal magnetic field
with reversing direction across the Galactic plane, while the vertical
field is directed in the same way above and below the plane (the right
hand plot in Fig.~\ref{f:widrow}).

%Halo magnetic fields can provide crucial constraints to dynamo
%models. Available numerical models (e.g.\ Ruzmaikin? Shukurov? Moss?
%Hanasz?  widrow other) predict various magnetic field structures out of
%the Galactic plane [CHECK!!!!], therefore observational constraints on
%the strength and direction of the magnetic field in the halo are
%crucial to distinguish between these models.

These two configurations can be observationally distinguished mainly
in two ways: (1) an even or odd azimuthal magnetic field direction
with respect to the Galactic plane; and (2) a symmetric or
antisymmetric vertical magnetic field configuration across the
Galactic plane. Below, observations trying to clarify these two
distinctions are briefly discussed.

\articlefigure{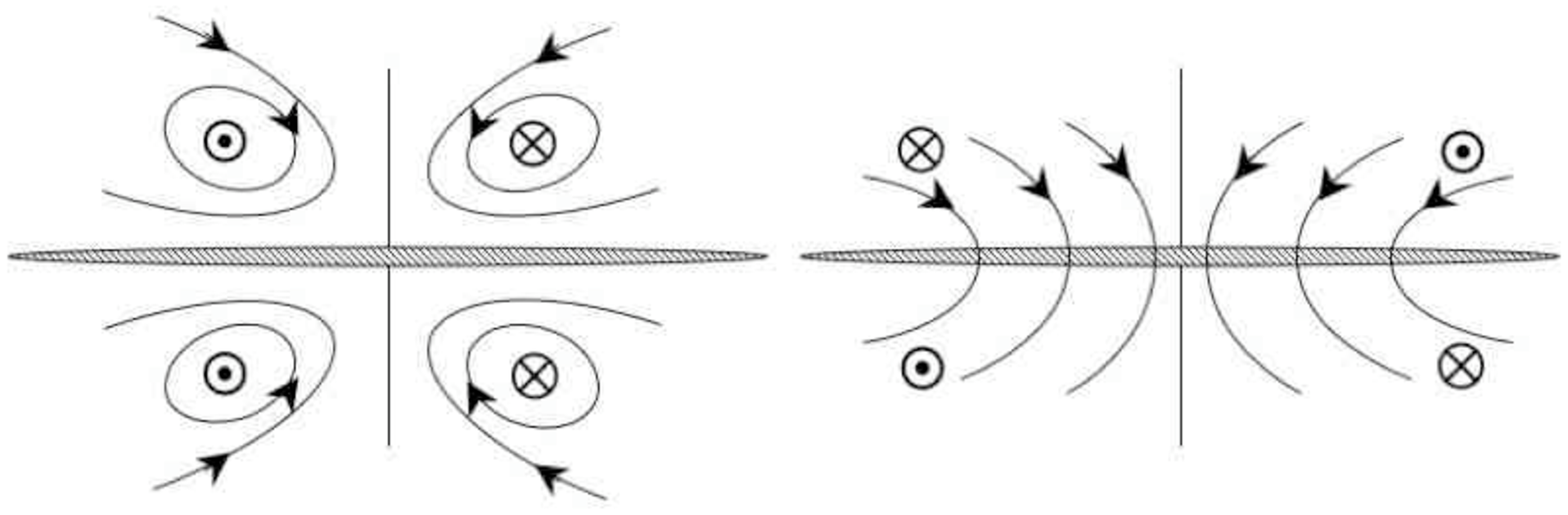}{f:widrow}{Possible large-scale
magnetic field configurations of the Milky Way. The viewer is located
outside the Galaxy in the plane, looking towards the Galactic center
(the center of each graph). Magnetic field towards the viewer is
denoted by a dot, field away from the observer as a cross.}

\paragraph{Direction of the azimuthal magnetic field component above and below the Galactic plane}

A large-scale 'butterfly pattern' in the azimuthal field component
parallel to the line of sight, as shown on the right in
Fig.~\ref{f:widrow}, is obvious in rotation measure data
\citep[e.g.][]{sk80}. However, this configuration exists only towards
the inner Galaxy. This has been interpreted as the signature of an A0
dynamo in the Milky Way by \citet{hmb97}. However, other authors have
pointed out that the quadrupolar structure in rotation measure could
also be caused by nearby, large, magnetized interstellar features such
as the North Polar Spur (Stil et al, in prep; Wolleben et al, in
prep).
		   
\paragraph{The vertical magnetic field component in the Galactic halo}

Even though our viewpoint at the Solar radius may obscure the clarity
of the configurations in Fig.~\ref{f:widrow} a bit, we can hope to
detect vertical magnetic field strengths using Faraday rotation of
extragalactic background sources to estimate a vertical component of
the Galactic magnetic field. The first ones to try this were
\citet{hq94}, who found a small vertical magnetic field of
$B=0.3\pm0.2~\mu$G from south to north. However, they worked with the
a priori assumption that the field was anti-symmetric and fitted
sources in north and south to one vertical magnetic field direction
only. This assumption was omitted by \citep{mgh10}, who obtained
rotation measures of more than 800 polarized extragalactic sources
towards both the north and the south Galactic poles at latitudes $|b|
> 70^{\circ}$. They concluded that there is no vertical magnetic field
component towards the north ($B_{vert} = 0.0 \pm 0.02~\mu$G), and a
small vertical component in the south ($B_{vert} = 0.31 \pm
0.03~\mu$G). This difference could be explained by smaller scale
magnetic field structure in one or both of the hemispheres, as also
shown from starlight polarization \citep{berdyuginteerikorpi01}.

\citet{tss09} used rotation measure values from NVSS sources across
the whole northern sky to infer vertical magnetic field components
$B_{vert} = 0.14\pm0.02~\mu$G in the north and $B_{vert} =
0.3\pm0.03~\mu$G in the south. Their southern values are consistent
with \citet{mgh10}. The difference in obtained $B_{vert}$ in the north
between the two studies may be attributed to the rotation measure of
the North Polar Spur, which was subtracted in the Mao et al.\ study
but not in Taylor et al.

\section{After the CGPS}
\label{s:future}

A bright future lies ahead for radio astronomy, and in particular for
radio polarimetry and cosmic magnetism. A number of new telescopes,
most of them path finders for the Square Kilometer Array, are coming
online in the next few years, in an era of much increased recognition
of the importance of magnetic fields in diffuse media. In combination
with state-of-the-art technology, the method of Rotation Measure
Synthesis has become feasible. RM Synthesis is based on a Fourier
transform of the complex polarization as a function of wavelength
squared, to obtain complex polarization as a function of Faraday depth
$\phi \propto \int_{x1}^{x2} n_e \mathbf{B}\cdot\mathbf{dl}$, where
$x1$ and $x2$ are locations along the line of sight. For polarized
radiation from a background synchrotron source Faraday rotated by a
foreground component, Faraday depth is equal to rotation measure.

The traditional method of determining RM from a small number of
frequencies, RM~$= \Delta\theta/\Delta\lambda^2$, becomes useless when
synchrotron emitting and Faraday rotating plasma are mixed. In this
case RM Synthesis provides a spectrum in Faraday depth, which details
the mixed (Faraday-thick) medium, and any other RM and emission
components along the line of sight. Although caution has to be
observed when using RM Synthesis with a small number of frequency
channels (Rudnick et al., in prep), the method constitutes a major
step ahead in magnetic field research through radio polarimetry.
For description and discussion of the method of RM Synthesis, see
\citet{b66, bb05a, bb05b, fss10}.

\subsection{On-going and future radio polarimetric surveys}

The largest radio polarimetric project currently underway, measured in
sky and frequency coverage, is the Galactic Magneto-Ionic Medium
Survey \citep[GMIMS, PI Wolleben,][]{wlc10}. GMIMS endeavors to survey
the whole sky from 300 to 1800 MHz continuously in 6 different surveys
with frequency coverages of 300 to 800~MHz, 800 to 1300~MHz and 1330
to 1800~MHz, in both Northern and Southern hemispheres. The Northern
1300-1800~MHz survey is currently being observed with the DRAO 26-m
single dish \citep{wfl10}. The Southern 1300-1800~MHz survey will be
filled in by the on-going associated Southern Twenty-centimeter
All-sky Polarization Survey (STAPS, PI Haverkorn), and the Southern
300-800~MHz survey has also started, both with the Parkes 64-m single
dish. At slightly higher frequencies than GMIMS, the S-band
Polarization All-Sky Survey (S-PASS, PI Carretti) from 2.2 to 2.4~GHz
has just been completed and analysis is ongoing. The Galactic Arecibo
L-band Feed Array Continuum Transit Survey \citep[GALFACTS, PI
Taylor,][]{ts10} has just started to survey the Arecibo sky
(declinations -1$^{\circ}$ to 38$^{\circ}$) in radio polarimetric mode
from 1225~MHz to 1525~MHz, adding higher resolution to the GMIMS
results.

A few more years in the future brings radio polarimetric surveys with
the SKA path finders. One of the Survey Science Programs of the
Australian SKA Path finder (ASKAP) under construction in Western
Australia is the POlarization Sky Survey of the Universe's Magnetism
(POSSUM, PIs Gaensler, Taylor, and Landecker), focusing on
high-resolution surveying of the Southern polarized sky in
L-band. POSSUM is proposed as a three-part survey: POSSUM-Wide,
POSSUM-Deep and POSSUM-Diffuse. POSSUM-Wide is an all-sky survey which
will observe 3 million polarized extragalactic point sources to obtain
an ``RM Grid'' \citep{bg04}. POSSUM-Deep is a deep observation of
30~square degrees on the sky to determine the polarization properties
of faint sources. Diffuse POSSUM intends to survey the whole sky with
the maximum frequency coverage of 700 to 1800~MHz to image diffuse
polarization with RM Synthesis. Depending on the outcome of the time
allocation process, the POSSUM surveys will be observed commensally
with other ASKAP surveys with similar observing constraints.

At low frequencies, the LOw Frequency ARray (LOFAR) is able to detect
low rotation measures and therefore weak magnetic fields, and is
therefore ideally suited to investigate magnetic fields in the
Galactic halo. A Galactic Science subgroup within the Cosmic Magnetism
Key Science Project has proposed broadband radio polarimetric
observations of many fields away from the galactic plane.

LOFAR consists of two kinds of antennas, the Low Band Antennas (LBAs)
with a frequency coverage of 10~-~90~MHz, and the High Band Antennas
(HBAs) from 110~-~240~MHz. These antennas are grouped in stations,
which are located in the Netherlands and throughout Europe. At this
time, international stations are being built in Germany, France,
Sweden, and the UK, although more international stations are in
various stages of planning. Each station consists of 96 LBA antennas,
and 48 (Dutch) or 96 (international) HBA antennas. A cabinet at each
station collects the signals from all antennas and performs some
processing such as station calibration or station beam
forming. High-speed glass-fiber connections connect all stations to
the Blue Gene P supercomputer at the Computing Centre of the
University of Groningen, where signals are processed through one of
the available data reduction pipelines. At the moment, pipelines for
imaging, transient detection, pulsars, high-energy cosmic ray air
showers, and rotation measure synthesis are in various stages of
development and testing.

\section{Conclusions}

The polarized Canadian Galactic Plane Survey has played a large role
to convince a growing part of the astronomical community of the
importance of Galactic magnetic fields in many physical processes. The
interactions of magnetic fields with the gaseous and cosmic ray
components in the Milky Way, as described above, can only be
understood if sufficient high-quality observations of magnetic field
strengths and directions and their influence on the surrounding ISM
are available. Here, the CGPS is in particular valuable because of its
multi-wavelength character, allowing both magnetic fields and ISM
components to be studied simultaneously, and because the CGPS team
excels in thorough and careful analysis of the observations to reach
solid observational conclusions.  Partially due to the CGPS
polarization efforts, galactic magnetism has become a thriving field
of research. Now, cosmic magnetism is one of five key science areas
for the Square Kilometre Array, assuring many exciting developments in
the years to come.

\acknowledgements The author would like to thank Rainer Beck, Katia
Ferri\`ere, Tyler Foster, and Naomi McClure-Griffiths for good advice
and enlightening discussions.

\bibliography{haverkorn_marijke}

\end{document}